\documentclass[aps,preprint,floatfix,10point]{revtex4}

\usepackage{graphicx,color}
\usepackage{psfrag}
\usepackage{amssymb}
\usepackage{eso-pic}
\usepackage[dvips,letterpaper]{geometry} 
\usepackage{amsmath}

\begin{document}

\definecolor{cbl}{rgb}{0,0,1}                
\newcommand{\cbl}[1]{\textcolor{cbl}{#1}} 

\definecolor{cbr}{rgb}{1,0,0}                
\newcommand{\cbr}[1]{\textcolor{cbl}{#1}} 

%
%
\def\oti{{\otimes}}
\def\lb{ \left[ }
\def\rb{ \right]  }
\def\tilde{\widetilde}
\def\bar{\overline}
\def\hat{\widehat}
\def\*{\star}
\def\[{\left[}
\def\]{\right]}
\def\({\left(}		\def\BL{\Bigr(}
\def\){\right)}		\def\BR{\Bigr)}
	\def\BBL{\lb}
	\def\BBR{\rb}
%
%
\def\zb{{\bar{z} }}
\def\zbar{{\bar{z} }}
\def\frac#1#2{{#1 \over #2}}
\def\inv#1{{1 \over #1}}
\def\half{{1 \over 2}}
\def\d{\partial}
\def\der#1{{\partial \over \partial #1}}
\def\dd#1#2{{\partial #1 \over \partial #2}}
\def\vev#1{\langle #1 \rangle}
\def\ket#1{ | #1 \rangle}
\def\rvac{\hbox{$\vert 0\rangle$}}
\def\lvac{\hbox{$\langle 0 \vert $}}
\def\2pi{\hbox{$2\pi i$}}
\def\e#1{{\rm e}^{^{\textstyle #1}}}
\def\grad#1{\,\nabla\!_{{#1}}\,}
\def\dsl{\raise.15ex\hbox{/}\kern-.57em\partial}
\def\Dsl{\,\raise.15ex\hbox{/}\mkern-.13.5mu D}
%
%
\def\ga{\gamma}		\def\Ga{\Gamma}
\def\be{\beta}
\def\al{\alpha}
\def\ep{\epsilon}
\def\vep{\varepsilon}
\def\la{\lambda}	\def\La{\Lambda}
\def\de{\delta}		\def\De{\Delta}
\def\om{\omega}		\def\Om{\Omega}
\def\sig{\sigma}	\def\Sig{\Sigma}
\def\vphi{\varphi}
%
%
\def\CA{{\cal A}}	\def\CB{{\cal B}}	\def\CC{{\cal C}}
\def\CD{{\cal D}}	\def\CE{{\cal E}}	\def\CF{{\cal F}}
\def\CG{{\cal G}}	\def\CH{{\cal H}}	\def\CI{{\cal J}}
\def\CJ{{\cal J}}	\def\CK{{\cal K}}	\def\CL{{\cal L}}
\def\CM{{\cal M}}	\def\CN{{\cal N}}	\def\CO{{\cal O}}
\def\CP{{\cal P}}	\def\CQ{{\cal Q}}	\def\CR{{\cal R}}
\def\CS{{\cal S}}	\def\CT{{\cal T}}	\def\CU{{\cal U}}
\def\CV{{\cal V}}	\def\CW{{\cal W}}	\def\CX{{\cal X}}
\def\CY{{\cal Y}}	\def\CZ{{\cal Z}}
\def\rvac{\hbox{$\vert 0\rangle$}}
\def\lvac{\hbox{$\langle 0 \vert $}}
\def\comm#1#2{ \BBL\ #1\ ,\ #2 \BBR }
\def\2pi{\hbox{$2\pi i$}}
\def\e#1{{\rm e}^{^{\textstyle #1}}}
\def\grad#1{\,\nabla\!_{{#1}}\,}
\def\dsl{\raise.15ex\hbox{/}\kern-.57em\partial}
\def\Dsl{\,\raise.15ex\hbox{/}\mkern-.13.5mu D}
%
%
\font\numbers=cmss12
\font\upright=cmu10 scaled\magstep1
\def\stroke{\vrule height8pt width0.4pt depth-0.1pt}
\def\topfleck{\vrule height8pt width0.5pt depth-5.9pt}
\def\botfleck{\vrule height2pt width0.5pt depth0.1pt}
\def\Zmath{\vcenter{\hbox{\numbers\rlap{\rlap{Z}\kern
0.8pt\topfleck}\kern 2.2pt
                   \rlap Z\kern 6pt\botfleck\kern 1pt}}}
\def\Qmath{\vcenter{\hbox{\upright\rlap{\rlap{Q}\kern
                   3.8pt\stroke}\phantom{Q}}}}
\def\Nmath{\vcenter{\hbox{\upright\rlap{I}\kern 1.7pt N}}}
\def\Cmath{\vcenter{\hbox{\upright\rlap{\rlap{C}\kern
                   3.8pt\stroke}\phantom{C}}}}
\def\Rmath{\vcenter{\hbox{\upright\rlap{I}\kern 1.7pt R}}}
\def\Z{\ifmmode\Zmath\else$\Zmath$\fi}
\def\Q{\ifmmode\Qmath\else$\Qmath$\fi}
\def\N{\ifmmode\Nmath\else$\Nmath$\fi}
\def\C{\ifmmode\Cmath\else$\Cmath$\fi}
\def\R{\ifmmode\Rmath\else$\Rmath$\fi}

\def\barray{\begin{eqnarray}}
\def\earray{\end{eqnarray}}
\def\beq{\begin{equation}}
\def\eeq{\end{equation}}

\def\n{\noindent}

\def\Tr{\rm Tr} 
\def\xvec{{\bf x}}
\def\kvec{{\bf k}}
\def\kvecp{{\bf k'}}
\def\omk{\om{\kvec}} 
\def\dk#1{\frac{d\kvec_{#1}}{(2\pi)^d}}
\def\dkvec{ \frac{d^3 \kvec}{(2\pi)^3}}
\def\2pid{(2\pi)^d}
\def\ket#1{|#1 \rangle}
\def\bra#1{\langle #1 |}
\def\vol{V}
\def\adag{a^\dagger}
\def\rme{{\rm e}}
\def\Im{{\rm Im}}
\def\pvec{{\bf p}}
\def\fermiS{\CS_F}
\def\cdag{c^\dagger}
\def\adag{a^\dagger}
\def\bdag{b^\dagger}
\def\vvec{{\bf v}}
\def\muhat{{\hat{\mu}}}
\def\vac{|0\rangle}
\def\pcut{{\Lambda_c}}
\def\chidot{\dot{\chi}}
\def\gradvec{\vec{\nabla}}
\def\psitilde{\tilde{\Psi}}
\def\psibar{\bar{\psi}}
\def\psidag{\psi^\dagger} 
\def\m{m_*}
\def\up{\uparrow}
\def\down{\downarrow}
\def\Qo{Q^{0}}
\def\vbar{\bar{v}}
\def\ubar{\bar{u}}
\def\smallhalf{{\textstyle \inv{2}}}
\def\smallsqrt{{\textstyle \inv{\sqrt{2}}}}
\def\rvec{{\bf r}}
\def\avec{{\bf a}}
\def\pivec{{\vec{\pi}}}
\def\svec{\vec{s}} 
\def\phivec{\vec{\phi}}
\def\daggerc{{\dagger_c}}
\def\Gfour{G^{(4)}}
\def\dim#1{\lbrack\!\lbrack #1 \rbrack\! \rbrack }
\def\qhat{{\hat{q}}}
\def\ghat{{\hat{g}}}
\def\nvec{{\vec{n}}}
\def\bull{$\bullet$}
\def\ghato{{\hat{g}_0}}
\def\r{r}
\def\deltaq{\delta_q}
\def\gcharge{g_q}
\def\gspin{g_s}
\def\deltas{\delta_s}
\def\gQC{g_{AF}} 
\def\ghatqc{\ghat_{AF}}
\def\xqc{x_{AF}}
\def\mhat{\hat{m}}
\def\xup{x_2}
\def\xdown{x_1}
\def\sigmavec{\vec{\sigma}}
\def\xopt{x_{\rm opt}}
\def\Lambdac{{\Lambda_c}}
\def\angstrom{{{\scriptstyle \circ} \atop A}     }
\def\AA{\leavevmode\setbox0=\hbox{h}\dimen0=\ht0 \advance\dimen0 by-1ex\rlap{
\raise.67\dimen0\hbox{\char'27}}A}
\def\ratio{\gamma}
\def\Phivec{{\vec{\Phi}}}
\def\singlet{\chi^- \chi^+} 
\def\mhat{{\hat{m}}}

\def\Im{{\rm Im}}
\def\Re{{\rm Re}}

\def\xstar{x_*}
\def\sech{{\rm sech}}
\def\Li{{\rm Li}}
\def\dim#1{{\rm dim}[#1]}
\def\ep{\epsilon}
\def\free{\CF}
\def\Fhat{\digamma}
\def\ftilde{\tilde{f}}
\def\muphys{\mu_{\rm phys}}
\def\xitilde{\tilde{\xi}}
\def\CI{\mathcal{I}}
\def\nhat{\hat{n}}
\def\ef{\epsilon_F}
\def\as{a_s}
\def\diffk{|\kvec - \kvec' |}
\def\bfT{{\bf T}}

\def\bfC{{\bf C}}
\def\bfP{{\bf P}}

\def\Ct{ \tilde{C} }
\def\Tt{ \tilde{T} }
\def\Pt{ \tilde{P} }
\def\etatilde{\tilde{\eta}}
\def\pt{\tilde{p}}
\def\Ttilde{\Tt}
\def\Ctilde{\Ct}
\def\Ptilde{\Pt}

\def\zbar{{\overline{z}}}
\def\dbar{{\overline{d}}}
\def\pbar{{\overline{\d }}}
\def\Abar{{\overline{A}}}

\def\Gev{{\rm Gev}}

\def\rhovac{\rho_{\rm vac}}

\def\Z{\mathbb{Z}}

\def\Lambdap{k_p}
\def\Ep{E_p} 
\def\Lambdac{k_c}

\def\psibar{\overline{\psi}}

\def\Principle{{\it Principle of Emptiness}}
\def\adot{\dot{a}}
\def\addot{\ddot{a}}
\def\adota{\frac{\adot}{a}}
\def\addota{\frac{\addot}{a}}
\def\rhovacphys{\rho_{\rm vac}}
\def\rhoLam{\rhovac}
\def\OmLam{\Omega_{\rm vac}}
\def\Index{\Delta N}
\def\rhovacCylinder{\rho_{\rm vac}^{\rm cyl}}
\def\minusIndex{ (N_f - N_b )}
\def\gee{g}
\def\rhorad{\rho_{\rm rad}}
\def\Omegarad{\Omega_{\rm rad}}
\def\rhodark{\rho_\Lambda}
\def\pvac{p_{\rm vac}}
\def\Ricci{\CR}
\def\rhoc{\rho_{\rm c}}
\def\rhovacbare{\rho_{\rm vac,0}}
\def\Omegam{\Omega_m}

\def\uk{u_\kvec}

\def\rhovachat{\rho_{\hat{\rm vac}}}
\def\pvachat{p_{\hat{\rm vac}}}
\def\ghat{\hat{g}}
\def\omegahat{\hat{\omega}}
\def\vac{| {\rm vac} \rangle}
\def\vachat{ | \hat{\rm vac} \rangle}
\def\Tmunu{T_{\mu\nu}}

\title{Scrutinizing the Cosmological Constant Problem\\ and a possible resolution}

\author{ Denis Bernard$^1$ and  Andr\'e  LeClair$^{1,2}$ }
\affiliation{ \centerline{$^1$Laboratoire de physique th\'eorique, Ecole Normale Sup\'erieure $\&$ CNRS, Paris,  France}\\
\centerline{$^2$ Physics Department, Cornell University,  Ithaca,  NY }\\ 
  }

\vskip 2.5 truecm

\begin{abstract}
We suggest a  new  perspective on the Cosmological Constant Problem  by  scrutinizing  its standard formulation. In classical and quantum mechanics without gravity,   there is no definition of the zero point of energy.    Furthermore,  the  Casimir effect only measures how the vacuum energy {\it changes}  as one varies a geometric modulus.     This leads us to propose that the physical  vacuum energy in a  Friedman-Lema\^itre-Robertson-Walker expanding universe only depends on the time variation of the scale factor $a(t)$.  Equivalently,  requiring that empty  Minkowski space is gravitationally stable is a principle that fixes the ambiguity in the zero point energy.  
On the other hand,  if there is a meaningful bare cosmological constant,  this prescription should be viewed as a fine-tuning.
  We describe two different choices of vacuum,   one of which is consistent with the current universe consisting only of matter and vacuum energy.  The resulting vacuum energy density $\rhovac$  is constant in time and  approximately   $k_c^2 H_0^2$, where $k_c$ is a momentum cut-off and $H_0$ is the current  Hubble constant;   for a cut-off close to the Planck scale,  values of $\rhovac$ in agreement with astrophysical measurements are obtained. 
Another choice of vacuum is more relevant to the early universe consisting of only radiation and  vacuum energy,  and we suggest  it as a possible  model of inflation.    
\end{abstract}

\maketitle

\section{Introduction}

The Cosmological Constant Problem  (CCP)  is  now regarded as a major crisis of modern theoretical physics.
For some reviews of the ``old''   CCP,  see  \cite{Weinberg,Nobbenhuis,Martin,Rugh}.   
The problem is that simple  estimates of the zero point energy, or vacuum energy,  of a single bosonic  quantum field  yield a huge value (the standard 
calculation is reviewed below).    In the past,  this  led many theorists  to suspect that it was zero,  perhaps due to a principle such as
supersymmetry.    The modern version of the crisis is that astrophysical measurements reveal a very small positive
 value\cite{Astro}\cite{convert}: 
 \beq
 \label{rhovacmeasured}
\rho_\Lambda = 0.7 \times 10^{-29}\,  {\rm g ~cm^{-3}} = 2.8 \times 10^{-47}  ~ {\rm Gev^4/\hbar^3 c^5}.
\eeq
This value is smaller than the naive expectation  by a factor of $10^{120}$.    This embarrassing discrepancy
suggests a conceptual rather than  computational error.   The main point  of this paper is to question whether the 
CCP as it is currently stated is actually properly formulated.    
As we will see,  our line of reasoning leads to  an estimate of the cosmological constant which is much more reasonable,  and  of the correct order of magnitude.

 Let us  begin by ignoring gravity and considering only quantum mechanics in Minkowski space.   
Wheeler and Feynman once estimated  that there is enough zero point energy in a teacup  to boil all the Earth's oceans.   
This has led to the fantasy of tapping this energy for useful purposes,  however most physicists do not take such
proposals very seriously,  and in light of the  purported seriousness of the  CCP,  one should wonder why.  
     In fact,  there is no principle in quantum mechanics that allows a proper definition of
the zero  of  energy:  as in classical mechanics,  one can only measure changes in energy,  i.e. all energies can be
shifted by a constant with no measurable consequences.    Similarly, the  rules of statistical mechanics tell us that probabilities of configurations are ratios of (conditioned) partition functions, and these are invariant if the partition functions are  multiplied by a common factor as induced by a global shift of the energies.
Based on his understanding of quantum electrodynamics and his own treatment of the Casimir effect,  
Schwinger once said \cite{Schwinger},  
``...the vacuum is not only the state of minimum energy,  it is the state of {\it zero} energy,  zero momentum,
zero angular momentum,  zero charge, zero whatever.''     One should not confuse zero point energy
with ``vacuum fluctuations''  which refer to loop corrections to physical processes:    
photons do not scatter off the vacuum energy,  otherwise they would be unable to traverse the universe.
      All of this  strongly suggests that it is impossible to harness vacuum energy in order to do work, which  in turn  calls into question whether it could be a source of gravitation.   

The Casimir effect is often correctly cited as proof of the reality of vacuum energy.   However it needs to  be emphasized that
what is actually measured is the {\it change }  of  the vacuum energy as one  varies a geometric modulus,  i.e.  how it depends on
this modulus,  and this is unaffected by an arbitrary shift of the zero of energy. 
  The classic experiment is to measure the force between two plates as one changes their separation;
the modulus in question here is the distance $\ell$ between the plates and the force depends on how the vacuum energy varies with this separation. 
The Casimir force $F(\ell)$ is minus the derivative of the electrodynamic vacuum energy $E_{\rm vac}(\ell)$  between the two plates, $F(\ell)=-d E_{\rm vac}(\ell)/ d\ell$.
An arbitrary shift of the vacuum energy by a constant  that is independent of $\ell$ does not affect the measurement.   
For the electromagnetic field,  with two polarizations,   the well-known result is that the 
energy density between the plates is $\rho_{\rm vac}^{\rm cas} =  - \pi^2/720 \ell^4$.  Note that this is 
an attractive force;   as we will see,  in the cosmic context a repulsive force requires an over-abundance of
fermions.    It is also clear that the Casimir effect is an infra-red phenomenon that has nothing to do with
Planck scale physics.   Our cosmological proposal will actually involve a mixing of the infra-red (IR) and
ultra-violet (UV).    

For reasons that will be clear,  let us illustrate the above remark on the Casimir effect  with another version of it:   the vacuum energy in the  finite size geometry of
a higher dimensional cylinder.   Namely,    consider a massless quantum bosonic field 
on a Euclidean space-time geometry of  $S^1 \otimes R^3$  where the circumference of the circle $S^1$ is $\beta$.    Viewing the compact direction as spatial,  the momenta  in that 
direction are quantized and the vacuum energy density is 
\beq
\label{cylinder}
\rhovacCylinder = \inv{2 \beta}  \sum_{n \in \Zmath}  \int   \frac{d^2 \kvec}{(2 \pi)^2}  \sqrt{ \kvec^2 + (2 \pi n/\beta)^2 }    
 = - \beta^{-4} \pi^{3/2}  \Gamma (-3/2) \zeta (-3) + {\rm const.}
\eeq
Due to the different boundary conditions in the periodic verses finite size directions,   
$\rho_{\rm vac}^{\rm cas} (\ell)  =   2 \rhovacCylinder (\beta = 2\ell)$,  where the overall factor of $2$ is 
because of the two photon polarizations.    
The above integral  is divergent,   however if one is only interested in its $\beta$-dependence,   it can be regularized using   the Riemann zeta function
giving the above expression.     Note that the (infinite)  constant that has been  discarded in the regularization  is  actually  at the origin
of the CCP.     What is measurable is the $\beta$ dependence.     One way to convince oneself  that this regularization is meaningful is to  view the
compactified direction as Euclidean time,  where now $\beta = 1/T$ is an inverse temperature.     The quantity $\rhovacCylinder$ is now the free energy density of a single scalar field,  and standard quantum statistical mechanics gives the convergent expression which is just the standard black-body formula:
\beq
\label{cylinder2}
\rhovacCylinder =  \inv{\beta}  \int    \frac{d^3 \kvec}{(2\pi)^3}  \log \( 1 - e^{-\beta k } \) =  - \beta^{-4}  \frac{\zeta (4)}{2 \pi^{3/2} \Gamma (3/2)}  = - \frac{\pi^2}{90}  T^4 .
\eeq
The two above expressions (\ref{cylinder}, \ref{cylinder2})  are equal due to a  non-trivial functional identity satisfied by the $\zeta$ function:
$\xi (\nu) = \xi (1-\nu)$  where $\xi (\nu) = \pi^{-\nu/2} \Gamma (\nu/2 ) \zeta (\nu)$.  (See for instance the appendix in  \cite{ALRiemann} in this context.)   
The comparison  of eqns. (\ref{cylinder},\ref{cylinder2}) strongly manifests 
the arbitrariness of the zero-point energy:  whereas there is a divergent constant in (\ref{cylinder}),  
from the point of view of quantum statistical mechanics,  the expression (\ref{cylinder2}) is actually 
convergent.
Either way of viewing the problem allows a shift of $\rhovacCylinder$ by an arbitrary constant with no measurable consequences.   
For instance, such a shift  would not affect thermodynamic quantities like  the entropy or density since they are derivatives of the free energy;   the only thing that is measurable is the $\beta$ dependence.

We now include gravity in the above discussion.    Before  stating  the  basic hypotheses of our study,   we begin with 
general motivating remarks.   All forms of energy should be considered as possible sources of gravitation,  including
the vacuum energy.     However,  if one accepts the above arguments that the zero of energy is not  absolutely definable in quantum mechanics,  and that only the dependence of the vacuum energy on geometric moduli including the space-time metric is physically measurable, it then remains unspecified  how to incorporate vacuum energy as a source of gravity.      One needs an additional principle to fix the ambiguity.

The above observations on the Casimir energy were instrumental   toward  formulating such a principle,  as we now describe. 
 The cosmological Friedmann-Lema\^itre-Robertson-Walker (FLRW)  metric  has no modulus corresponding to a finite size analogous to $\beta$,  however it does have a  time dependent scale factor $a(t)$:
\beq
\label{FRW}
ds^2 =  g_{\mu\nu} dx^\mu dx^\nu =  - dt^2  +  a(t)^2 d\xvec \cdot d\xvec .
\eeq  
(We assume the spatial curvature $k=0$,   as shown by recent astrophysical measurements.)
  When $a(t)$ is constant in time,  the FLRW metric is just the Minkowski spacetime  metric.      
  This leads us  to propose that the dependence of the vacuum energy on 
the time variation of $a(t)$ is all that is physically meaningful,  in analogy with the $\beta$ dependence of $\rhovacCylinder$. 
This idea is stated as a principle below,  in terms of the stability of empty Minkowski space,  and is at the foundation of our conclusions.  

Let us quickly review the standard cosmology.      The Einstein equations are 
\beq
\label{Einstein}
G_{\mu\nu} \equiv R_{\mu\nu}  - \inv{2}  g_{\mu \nu}  \Ricci =  8 \pi G  T_{\mu\nu} 
\eeq
where  $G$ is  Newton's constant.
The stress energy tensor $T_{\mu\nu} = {\rm diag} (\rho,  p, p, p)$ where $\rho$ is the energy density and $p$ the pressure.  
The non-zero elements of the Ricci tensor are $R_{00} =  - 3 \,\ddot a/a$, $R_{ij} = ( 2 \adot^2 + a \addot ) \delta_{ij}$, and the Ricci scalar is $\Ricci =g^{\mu\nu} R_{\mu\nu} = 6 \( (\dot a/a)^2 + \ddot a/a \)$, where over-dots  refer to time-derivatives. 
The  temporal and spatial Einstein equations (\ref{Einstein})  for the  FLRW metric are then the  Friedmann equations: 
\barray
\label{Friedmann1}
\( \adota \)^2 &=&  \frac{8 \pi G}{3}  \rho ,  
\\ 
\label{Friedmann2}
\( \adota\)^2 + 2\, \addota & =& - 8 \pi G  \, p  .
\earray
Taking a time derivative of the first equation and using the second,  one obtains
\beq
\label{energyconservation}
\dot{\rho} = - 3 \( \adota \) (\rho + p), 
\eeq 
which expresses the usual   energy conservation.   The above three equations are thus not functionally independent,  
the reason being that Bianchi identities relate the two Friedmann equations  to the energy  conservation equation (\ref{energyconservation}). 
 The total energy  density is usually assumed to consist of a
mixture of three non-interacting fluids,   radiation,  matter, and dark energy,  $\rho = \rhorad + \rho_m  + \rhodark$,    each of which satisfies eq. (\ref{energyconservation}) 
separately,  with $p=w\rho$ for $w=1/3, \, 0$ and $-1$ respectively.    
 Then,  eq.(\ref{energyconservation})  consistently implies
$\dot{\rho}_\Lambda  =0$.     The energy density is related to the classical cosmological constant 
as  $\Lambda = 8 \pi G \rho_\Lambda$.

In this paper we will assume that dark energy comes entirely from vacuum energy,  $\rhodark = \rhovac$.     
The vacuum energy $\rhovac$ is  a  quantum expectation value, 
\beq
\label{rhovacdef}
\rhovac = \langle  \CH  \rangle  =  \langle {\rm vac}  |  \CH  \vac ,
\eeq
where $\CH$ a quantum operator corresponding to the energy density,  which is usually associated with $T_{00}$.    

Apart from the ambiguity of the zero point energy,  several other points should be emphasized.     We will be studying the semi-classical Einstein equations, 
where  on the right hand side we include  the contribution of vacuum energy
$\langle  T_{\mu\nu} \rangle  =  \langle {\rm vac} |  T_{\mu\nu}  | {\rm vac} \rangle$  for some
choice of vacuum state $\vac$.    Given the very low energy scale of expansion in the current universe, 
and the weakness of cosmological gravitational fields,  
it is very reasonable to assume that there is no need to quantize the gravitational field itself in the present epoch. 
One  hypothesis of the standard formulation of the CCP is that the vacuum stress tensor is proportional to the
 metric\cite{Weinberg}.  
In an expanding universe,   the hamiltonian is effectively time dependent, and there is not necessarily a unique choice of $\vac$, 
 and, in contrast  to flat Minkowski space, no Lorentz symmetry argument \cite{footnote2} 
enforces that $\vev{T_{\mu\nu}}\propto g_{\mu\nu}$.   One needs extra information that characterizes $\vac$.   
This implies that  $\langle T_{\mu\nu} \rangle$  is not universal since it depends on $\vac$,  and thus,  for example,  cannot always be expressed in terms of purely geometric properties with no reference to the data of $\vac$.   
One  mathematical consistency condition is  $D^\mu \langle T_{\mu \nu} \rangle  = 0$ if the various components of
the total energy are separately conserved,   where $D^\mu$ is the covariant
derivative,  which is the statement of energy conservation.    However this may not 
follow from  $D^\mu T_{\mu\nu} =0$  since $\vac$ may be time dependent.    Also,   $\langle T_{\mu \nu} \rangle $   
is not necessarily 
  expressed in terms of manifestly convariantly conserved tensors such
as $G_{\mu \nu}$,  again because it depends on $\vac$.   In fact,  the only convariantly conserved geometric tensor that is second order in time derivatives is $G_{\mu\nu}$, and    if $\langle \Tmunu \rangle  \propto G_{\mu\nu}$,   
this would just amount to a renormalization of Newton's constant $G$.   
  
The second point is that if one includes $\langle T_{\mu \nu} \rangle$ as a source  in Einstein's equations,  then since 
it depends on $a(t)$ and its time derivatives,   doing so can be thought of as studying the back-reaction of this
vacuum energy on the geometry.    The resulting equations must be solved self-consistently and there is no
guarantee that there is a solution consistent with energy-conservation.

Having made these  preliminary observations,     let us state all of the hypotheses that this work is based upon,    
which specify either the vacuum states or the nature of their stress-tensor. They are the following: 

[i]   As a criterion  to identify possible vacuum states $\vac$,   we look for preferred quantization schemes such that
  $\vac$ is an eigenstate of the hamiltonian at all times,  which     
   implies there is no particle production.  

[ii]    We calculate  a bare  $\rho_{\rm vac, 0}$  from the hamiltonian,  i.e.  $\rho_{\rm vac,0} =  \langle {\rm vac} |  \CH  \vac$  where $\CH$ is the quantum  hamiltonian energy density operator.
The calculation is regularized with a sharp cut-off $k_c$ in momentum space in order to make contact with the usual statement of the CCP. 

[iii]    We propose the  following  principle which prescribes how to define a physical $\rhovac$ from $\rho_{vac,0}$:   
{\it  Minkowski space that is empty of matter and radiation should be stable, that is, static.}    This requires that the physical
$\rhovac$ equal zero when $a(t)$ is constant in time.     This leads to a $\rhovac$ that depends on $a(t)$ and its derivatives,  and also the cut-off.      

[iv]  Given this $\rhovac$,      we assume the components of the vacuum  stress energy tensor have the form of a cosmological constant:   
\beq
\label{Tuu} 
\langle T_{\mu \nu} \rangle   =    -\rhovac  \,  g_{\mu\nu} .
\eeq
We provide some support for this hypothesis in section III,  where we compare our calculation with 
manifestly covariant calculations performed in the past\cite{BunchDavies}.    We are going to check the consistency
of this assumption in the next point [v].  

[v]     We include  $\langle T_{\mu\nu} \rangle$   in Einstein's equations and solve them self-consistently,   assuming that vacuum energy and other forms of energy are separately conserved.      In other words we study the
consistency of the back-reaction of the vacuum energy on the geometry.     The consistency condition is $\dot{\rho}_{\rm vac} =0$, which is equivalent to $D^\mu  \langle T_{\mu\nu} \rangle  =0$.    There is no guarantee there is such a solution since $\rhovac$ depends on $a(t)$ and its derivatives.

Certainly one may question the validity of these assumptions.   However in our opinion,   they are rather conservative in that they do not invoke symmetries,  particles, nor other,  perhaps higher dimensional  structures, that are not yet  known to exist.  The purpose of this paper is to work out the logical consequences of these modest hypotheses.   
Our main findings are the following:

$\bullet$~   If there is a  cut-off in momentum space $k_c$,   then by dimensional
analysis the vacuum energy density has symbolically  the  ``adiabatic''   expansion (up to constants):
\beq
\label{expansion}
\rho_{\rm vac,0} =  k_c^4   +  k_c^2  \,  \hat{R}     +  \hat{R}^2   + \cdots  
\eeq
where  $\hat{R} $ is related to the curvature and is a linear combination of  $(\adot /a )^2$ and $\addot/a$, depending on the choice of $\vac$. 
The principle of the stability of empty Minkowski space [iii]   leads us to discard the $k_c^4$ term,   but not the other terms since they depend on time derivatives of $a(t)$.     The vacuum energy is now viewed as 
a {\it low energy}  phenomenon,  like the Casimir effect.   
Other regularization schemes,  based for example on point-splitting\cite{BunchDavies},   insist on a finite $\rhovac$ and thus  discard the first two terms.     According to our principles,   the second term must be kept since it depends dynamically on the geometry.       
In the current universe $\hat{R}$  it is approximately on the order of $ H_0^2$,  where  $H_0$ is the Hubble constant,   and if the cut-off $k_c $ is on the order of the Planck energy,  then the resulting value of $\rhovac$ is the right order of magnitude in comparison with the measured value (\ref{rhovacmeasured}), namely 
$\rhovac  \approx  (k_p H_0)^2 = 3.2 \times 10^{-46} \,  \Gev^4,$
using for $H_0$ the present value of the Hubble constant.
 The $\hat{R}^2$  is much too small to explain the measured value.  
We emphasize that our $\rhovac$ is not simply proportional to $H^2$, 
 see eqns.  (\ref{rhovacphys},\ref{rhovachat}) below,    and is in fact constant in time for the self-consistent
 solutions that we find.   
There is nothing special about $H_0$ here,  since $\rhovac$ is constant in time;   we are simply evaluating 
it at the present time which involves $H_0$. 
A practical  point of view is that astrophysical observations  are telling us  that the 
$k_c^4$ term should be shifted away.   
More importantly,  it remains to determine whether the term that we do keep, $k_c^2 \hat{R}$,  
has physical consequences in agreement with observations,  which is the main purpose of our study.

\def\rhovacConst{\rho_{\rm vac,0}}

Is shifting away the $k_c^4$ term a fine-tuning?     
Let us address this in the context of the ADM/AD framework\cite{ADM,AD}.     
In the latter work it was shown that in classical general relativity,   once the value of the 
cosmological constant $\Lambda$ is fixed,  there is a unique choice of energy that is conserved,  i.e. there
is no more freedom to shift it.   
For
asymptotically flat
spacetimes,  this energy 
 was proven to be positive and only zero for Minkowski space\cite{Schoen,witten}.  Its 
``main importance is that it is related to the stability of Minkowski space as the ground state of general relativity'', to quote ref.\cite{witten}.
For de Sitter space, 
there are similar statements, though with some restrictions\cite{AD}.   
Let us suppose that for some as yet unknown physical reason,  perhaps due to quantum gravity,   there is a meaningful bare cosmological constant $\Lambda_0$.     Then the $\rhovacConst$ that we 
calculate  leads to an effective cosmological constant   $\Lambda_{\rm eff} =  \Lambda_0 +  8 \pi G \rhovacConst$ in semi-classical gravity,  and this is what  is actually measured.  
Our prescription [iii]  amounts to setting $\Lambda_{\rm eff}$ to zero in  Minkowski space where $\adot$ vanishes, by adjusting $\Lambda_0$ or the manifestly constant part of $\rhovacConst$.   
This  is  a fine-tuning if 
$\Lambda_0$ is unambiguously defined.      Our work has nothing more to say about this issue,  rather, 
the main point of this work is to study the effects and consistency of what remains,  i.e. the
 $ \Lambda_{\rm eff}  \sim k_c^2 \hat{R}$ term which does not vanish in an expanding geometry.   
One may argue that,  as in any field theory,  the parameters in the effective lagrangian must ultimately be chosen 
to match experiments.     
As we will see,   under certain conditions,   this term can mimic a cosmological constant  in a way that is consistent with the current
era of cosmology.

The analogy with the Casimir effect is clear both mathematically (compare eqn.  (\ref{cylinder}) and 
eqn. (\ref{rhovacbare}) below),
and physically.  In the Casimir effect,  as one pulls apart the plates in a controlled manner in an experiment, this induces a measurable force.
In cosmology the analog of the growing separation of the plates  is the expansion itself,  which induces an acceleration; 
the complication is that the effect of this back-reaction must be solved self-consistently,  as we will do.  
 Note that whereas the Casimir force is attractive,  to describe the positive accelerated 
expansion of the universe,  one needs a positive $\rhovac$,  which as we will explain,  requires an overabundance of fermions.

$\bullet$~   For a universe consisting of only matter and vacuum energy,  such as the present universe,  there is a choice of $\vac$ with the above $\rhovac$  that leads to a consistent solution if a specific relation between $k_c$ and the Newton constant $G$ is satisfied.  By ``consistent'',  we mean $\dot{\rho}_{\rm vac} =0$.   Our solution
for $a(t)$ is consistent with present day astrophysical observations if one ignores the very small 
radiation component. 
 In fact, as  we will show,  our solution $a(t)$,  eq. (\ref{asolMatter})  below,  once one matches the integration
 constants to their present measured values,     
is  identical to  the standard $\Lambda{\rm CDM}$  model of the present universe,  i.e. a universe consisting only
of a cosmological constant plus cold dark matter,  and is thus not ruled out by observations up to fairly large 
redshift $z < 1000$;  it certainly agrees with supernova observations,  which are  at low $z <2$.  To our knowledge this choice for $\vac$ has not been considered before.   Below,  we also remark on the cosmic coincidence problem in light of our 
result.    We speculate that the relation between $G$ and $k_c$ suggests that gravity itself arises from quantum fluctuations,  and we provide an argument that ``derives'' gravity from quantum mechanics.

$\bullet~$  For a universe consisting  of only radiation and vacuum energy,   there is another  {\it different}  choice of vacuum,  $\vachat$,    that also has a consistent solution, again only for a certain relation between $k_c$ and $G$.   
This vacuum has been studied before and is referred to as the conformal vacuum in the literature.     We suggest that this solution possibly  describes inflation,   without invoking an inflaton field,  and speculate on 
 a scenario to resolve the  ``graceful exit''  problem.        
We also argue  that when $H = \adot /a$ is large,   the first Friedmann  equation sets the scale  $H \sim k_c$,  which is the right 
order of magnitude if $k_c$ is the Planck scale.

It is worthwhile comparing our model with other, similar proposals.
  Based on  ``wave-function of the universe''  arguments\cite{Strominger},  or simply dimensional analysis\cite{Wu}, 
  it was proposed that $\rhovac  \sim  (k_p/d_H (t) )^2$,   where 
 $d_H$ is a dimensionful scale factor related to   the cosmological horizon;   roughly $d_H \sim a(t) t $.     In the present universe $d_H (t_0) \approx  t_0 \approx 1/H_0$,  
so this $\rhovac$ is also of the right magnitude.   The problem with it is that it is time-dependent,  
and ruled out by observations.   Different arguments based on unimodular gravity\cite{Ng,Dodelson} 
 also led to the 
proposal that $\Lambda \sim  1/d_H^2$.       The work that is closest to ours is 
by Maggiore and collaborators\cite{Maggiore}.   Our approach differs from all the above in
that our $\rhovac$ is constant in time,  in agreement with observations.

 The next two sections simply describe these two choices of vacua and analyze the self-consistency of the back-reaction. Our analysis is done using an adiabatic expansion. In the Conclusion,   we further discuss our results.

\section{Vacuum energy plus matter}

\subsection{Choice of vacuum and its energy density}

We first review the  standard version of the Cosmological Constant Problem.   
Since a free quantum field is an infinite collection of harmonic oscillators for each wave-vector $\kvec$,  we first review simple quantum mechanical versions  in order to point out the difference between bosons and fermions.   
 Canonical quantization of a bosonic mode \cite{zero1} of frequency  $\omega$ yields to a pair of creation and annihilation operators, $a,\ a^\dag$, with $[a, a^\dagger] = 1$,  and a hamiltonian $H=\frac{\omega}{2} ( a a^\dagger + a^\dagger a)  =    \omega ( a^\dagger a + \inv{2} ) $. The boson zero-point energy is thus identified as $\omega/2$.  For fermions,  the zero point energy has the opposite sign. Fermionic canonical quantization \cite{zero2} yields to grassmanian operators $b,\ b^\dag$, with $\{b, b^\dagger\} = 1,  b^2 = {b^\dagger}^2 =0$,  and a hamiltonian 
$H = \frac{\omega}{2} \(  b^\dagger b - b b^\dagger \) =  \omega ( b^\dagger b - \inv{2} )$.
The fermion zero-point energy is $-\omega/2$.  

In a free relativistic quantum field theory  with particles of mass $m$ in $3$ spatial dimensions,  the above applies with  $\omega_\kvec = \sqrt{\kvec^2 + m^2}$, where $\kvec$ is a $3$-dimensional wave-vector.    
Thus the zero-point vacuum energy density  is 
\beq
\label{rhovac}
\rhovac = \frac{N_b - N_f}{2}  \int  \dkvec   ~ \sqrt{\kvec^2 + m^2} ,
\eeq
 where $N_{b,f}$ is the number of bosonic, fermionic  particle species. 
Regularizing the integral with an ultra-violet cut-off $\Lambdac$ much larger  than $m$,   one finds
$\rhovac  \approx  (N_b - N_f)   \Lambdac^4/16 \pi^2$.    If $\Lambdac$ is taken to be the Planck energy $\Lambdap$,  then $\Lambdac^4 /16 \pi^2 =  10^{75}  ~\Gev^4$.
The modern version of the Cosmological Constant problem is the fact that this is too large by a factor of $10^{122}$ in comparison with the measured value.  
One should also note that  in the above calculation,   a positive value for $\rhovac$ requires more bosons than fermions,  contrary to the currently known particle content of the Standard Model.  

As explained in the Introduction,  we are interested in the vacuum energy 
of a free quantum field in the  non-static  FLRW background spacetime geometry. 
For simplicity we consider a single scalar field,  with action \cite{footnote3}
\beq
\label{Phiaction}
S =  \int dt \, d^3 \xvec  \,  \sqrt{|g|} \(-  \inv{2}   g^{\mu\nu}  \d_\mu \Phi  \d_\nu \Phi  -  \frac{m^2}{2}  \Phi^2 \) .
\eeq
In order to simplify the explicit time dependence of the action,  and thereby simplify the quantization procedure,   define a new field $\chi$  as    $\Phi = \chi/a^{3/2}$.    Then the action (\ref{Phiaction}),  after an integration by parts,  becomes:
\beq
\label{chiAction}
S = \int dt \,  d^3 \xvec \,  \inv{2}  \(   \d_t \chi   \d_t \chi  - \inv{a^2}  \gradvec \chi \cdot \gradvec \chi  -  m^2   \chi^2  +  
\CA (t)    \chi^2 \) ,
\eeq
where 
\beq
\label{Adef} 
\CA   \equiv  \frac{3}{4} \(\(  \adota \)^2  + 2 \addota \) .
\eeq
The advantage of quantizing $\chi$ rather than $\Phi$  is that most of the time  dependence is now in $\CA$,   so that there is no spurious time dependence in
the canonical momenta,  etc.   
The field can be expanded in modes:
\beq
\label{Chimodes}
\chi = \int  \frac{d^3 \kvec}{(2\pi)^{3/2} }  \(  a_\kvec  \, u_\kvec \,  e^{i \kvec \cdot \xvec} 
+  a^\dagger_\kvec  u_\kvec^*  \,  e^{-i \kvec \cdot \xvec }  \) , 
\eeq
where  the $a_\kvec $'s  satisfy canonical commutation relations 
$[ a_\kvec ,  a^\dagger_{\kvec'} ]  =  \delta (\kvec - \kvec')$.   The function $u_\kvec$ is time dependent and required to 
satisfy
\beq
\label{uEq}
(\d_t^2 +  \omega_\kvec^2  )  u_\kvec = 0,      ~~~~\omega_\kvec^2    \equiv (\kvec/a)^2  + m^2 -\CA .
\eeq 

The solution is the formal expression 
\beq
\label{uSolution}
u_\kvec =  \frac{1}{ \sqrt{2 W} } \,  \exp \(  i \int^t  W (s)  ds \),
\eeq
where $W$ satisfies the differential equation:
\beq
\label{Wdiff}
W^2  =  \omega_\kvec^2 + \frac{3}{4} ( \dot{W}/W  )^2  - \inv{2} \ddot{W}/ W
\eeq   
Let us assume that the time  dependence is slowly varying,  i.e. we make an adiabatic expansion.  
The above equation can be solved iteratively,  where to lowest approximation,  $W$ is the
above expression with $W$ replaced by $\omega_\kvec$ on the right hand side  of the differential equation.
In other words,  the ``adiabatic condition'' is   $\dot{\omega_\kvec}/\omega_\kvec \ll \omega_\kvec$.  

 As we now explain,  it appears one has to distinguish between massive verses massless particles.
 Consider for instance the term proportional to 
 $(\dot{\omega}_\kvec / \omega_\kvec )^2  =   (\adot/a)^2 \( \kvec^2 /(\kvec^2 + m^2 a^2) \)^2$. 
    When $m=0$  this gives a term which modifies $\cal{A}$,  as does the $\ddot{\omega}/\omega$ term.
    The adiabatic condition is simply $\adot/a \ll k$.    
    The result is that the   additional two terms on the right hand side of eq. (\ref{Wdiff})  
 (with $W=\omega_\kvec$),   give $W^2  = ( \kvec /a)^2  - \CR/6$,  i.e. $\CA$ is converted to $\CR/6$.   This dependence on the 
 Ricci scalar $\CR$ can be derived more directly using conformal time,  as in the next section.   
 In this section we will only be considering cosmological matter plus vacuum energy.   When 
 $m \neq 0$,  the additional terms do not simply convert $\CA$ to $\CR/6$.  
 In order to implement an adiabatic expansion in this case, we consider the opposite limit of $m$ large.  
  One way to perhaps justify this is  as follows.   We will ultimately be interested in this vacuum energy 
  in the presence of a non-zero density of real matter.   
 In 
    cosmology,
 ``matter'' refers to non-relativistic particles, 
and formally,  the non-relativistic limit 
 corresponds to $m\to \infty$,  e.g. $\sqrt{\kvec^2 + m^2 }  \approx m^2 + \kvec^2 /2m$.   
 More importantly,  matter is defined as having zero pressure.   For a relativistic fluid, 
 the contribution of each mode $\kvec$ to the pressure is  $p = n_\kvec    \kvec^2 /3 \omega_\kvec$ where
 $n_\kvec $ is the density.    One then sees that zero pressure corresponds to $m\to \infty$.   
 Here the  adiabatic condition is $\adot/a \ll (\kvec^2 + m^2 )^{3/2}/\kvec^2$ which is automatically satisfied
 in this limit.   
 In the limit $m \to \infty$,  the additional terms on the right hand side of  eq. (\ref{Wdiff}) actually vanish. 
  Thus,  to lowest order we simply take $W = \omega_\kvec$,  and to this order $ \dot{u}_\kvec =  i\, \omega_\kvec  u_\kvec$.     As we will show in the next sub-section,  
  for a  pressure-less fluid this  has a self-consistent back-reaction.  
  
With this choice of $u_\kvec$, 
and to lowest order in the adiabatic expansion,
 the  hamiltonian takes the standard form:
\beq
\label{Ham}
H = \inv{2}  \int d^3 \xvec   \(   \dot{\chi}^2  +  \inv{a^2}(\gradvec \chi)^2  + (m^2  -  \CA)  \chi^2  \) 
=  \inv{2}  \int  d^3 \kvec  ~ \omega_\kvec \(  a_\kvec^\dagger a_\kvec  +  a_\kvec a^\dagger_\kvec \)  .
\eeq
Importantly,   there are no  $a_\kvec^\dagger  a_{-\kvec}^\dagger$ terms,  which implies  the vacuum $\vac$ defined by
$a_\kvec \vac =0$  is an eigenstate of $H$ for all times,  i.e. there is no particle production,
 again to lowest order in the adiabatic expansion.   
By the translational invariance of the vacuum,  for the bare vacuum energy we finally have:
\beq
\label{rhovacbare}
\rhovacbare =     \inv{V}  \langle H \rangle = \inv{2}  \int  \dkvec   ~ \sqrt{ \kvec^2  + m^2   -  \CA   } ,
\eeq
where $V$ is the volume and we have used $\delta_\kvec (0) = V/(2\pi)^3$.  
  In obtaining  the above expression we have properly scaled by redshift factors:   $V\to a^3 V$,   the cut-off was scaled to $k_c /a$,  and we made the change of variables $\kvec \to a \kvec$.
Comparing the above equation with eq. (\ref{cylinder}),  the analogy with the Casimir effect is clear.

Introducing an ultra-violet cut-off $\Lambdac  $ as before,     one finds 
\beq
\label{rhovacalpha}
\rhovacbare  \approx     \frac{\Lambdac^4}{16 \pi^2 }  \[   1 + \alpha  +  \frac{\alpha^2}{8}  \(  1 + 2 \log (\alpha/4) \) \] ,
\eeq
where $\alpha \equiv  (m^2  -\CA)/ \Lambdac^2$ is assumed to be small and positive.
   Assuming that masses $m$ are all much smaller than the cut-off,
we approximate the above expression as:  
\beq
\label{rhovacapprox}
\rhovacbare  \approx     \frac{\Lambdac^4}{16 \pi^2 }\[  1  + \frac{m^2}{\Lambdac^2}
- \frac{\CA}{\Lambdac^2} + \frac{\CA^2}{8 \Lambdac^4} \] ,
\eeq
where we have neglected the logarithmic contribution. It should also be noticed that the  $\CA^2$ term  is beyond the  lowest order in the adiabatic expansion.

Now we apply the principle [iii] of the Introduction.    In empty Minkowski space,  by definition $\adot = \addot =0$  and  $\rhovac$ must be zero otherwise empty 
Minkowski spacetime  would not be static due to gravity.     Thus,   $\rhovacbare$ must be regularized to a physical  $\rhovac$ by subtracting the first two constant terms in brackets:
\beq
\label{rhovacphys}
\rhovac   \approx   \Index \[    \frac{ \Lambdac^2}{16\pi^2}   \CA   
-   \inv{128 \pi^2}  \CA^2  \] .
\eeq
such that $\rhovacphys =0$  when $\adot = \addot =0$.   Above,  we have included multiple species 
 $\Index = N_f - N_b$ where $N_{f,b}$ are the numbers of species of fermions,bosons.   
It is important to observe that  in the  cylindrical  version of the Casimir effect, eq. (\ref{cylinder}),  the analog of the first term above is proportional to $\zeta (-2)  k_c/\beta^3= 0$,  so that there is no analog of it in the Casimir effect.

Before proceeding,  let us first check that the above expression gives reasonable values.   
 In the present universe,   $ \adot /a = H_0 = 1.5 \times 10^{-42} \, \Gev$ is the Hubble constant,  and
$(\adot/a)^2 \approx  \addot/a$.     If $\Lambdac$ is taken to be the Planck energy $k_p$,   then  
$\rhovac  \sim  (k_p H_0)^2 = 3.2 \times 10^{-46} \,  \Gev^4$,
which at least is in the ballpark.    
Fortunately there are more fermions than bosons in the Standard Model of particle physics so that the above expression is positive.  
Each quark/anti-quark has two spin states, and comes in 3 chromodynamic colors.
The electron/positron has two spin states,  whereas a neutrino has one.   For 3  flavor generations,  this gives $N_f = 90$.   
Each  massless gauge boson has two polarizations, $8$ for QCD,  and $4$ for the electro-weak theory, 
  which leads to $N_f - N_b =  60$ including the 4 Higgs fields before spontaneous 
electro-weak symmetry breaking and the two graviton polarizations.   
Incidentally,  for  one generation $N_f = N_b = 30$,  so in order for  the  cosmological constant to be positive,
one needs at least 2 generations.
The measured value of the vacuum energy can be accounted for with a cut-off about an order of magnitude  below the Planck energy \cite{footnote5}, $\Lambdac  \approx 3  \times 10^{18}$ \Gev.  
We have ignored interactions  which modify the value of $\rhovac$,   however we  expect that they do not drastically change our results. One should also bear in mind that the sharp cutoff $k_c$ is meant to represent a cross-over from the effective theory valid at energy scale well below $k_c$ to that (including gravity) valid above $k_c$.

\subsection{Consistent backreaction}

Let us suppose that the only form of vacuum energy is  $\rhovacphys$ of the last section eq. (\ref{rhovacphys}),  and that $a(t)$ is varying slowly enough in time that the $\CA^2$ term can be neglected. 
Define  the dimensionless constant:
\beq
\label{GLambda}
\gee  =     \frac{ 3 \Index }{8 \pi }  G \, k_c^2
\eeq
such that 
\beq
\label{rhovacg}
\rhovac =   \frac{g}{6 \pi G}  \, \CA .
\eeq
Including $\rhovac$  in the total $\rho$,  the first Friedmann equation can be written as 
\beq
\label{Fried1}
\( 1-\frac{g}{3}  \) \( \adota \)^2  -  \frac{ 2g}{3}\,    \addota  =  \frac{8 \pi G}{3} \(  \rho_m  + \rhorad \)
\eeq
We emphasize that we have not modified the Friedman equation;  the extra terms on the left-hand side
come from $\rhovac$ which were originally on the  right-hand side of the first Friedmann  equation.  

As we now argue,    there is only a consistent solution when $g=1$.     
First consider the case where there is no radiation nor matter.     
Then eq. (\ref{Fried1}) implies $\ddot a/a =(3-g)(\dot a/a)^2/2g$.
Firstly,  note that this implies a constant expansion,  i.e. de Sitter space,   only if $g=1$.  
Secondly,  the pressure can  then be found from eq.  (\ref{Friedmann2})
\beq
\label{pvac}
 \pvac = -  \inv{g}  \,  \rhovac .
\eeq
Thus,  the equation of state parameter $w=-1/g$ when there is only $\rhovac$.  
However energy conservation requires  $\dot{\rho}_{\rm vac} =0$,  which  requires  $\pvac = - \rhovac$, i.e.   $g=1$.      
The solution is    $a(t) \propto  e^{Ht}$ for an arbitrary constant $H$,   and $\rhovac$ is independent of time,  as a cosmological constant must.    
 
\def\tin{t_i}

What is not immediately obvious is that a consistent solution can also be found when matter  is included,  again 
when $g=1$.       
At the current time  $t_0$,  as usual define the critical density  $\rho_c =  3 H_0^2 /8 \pi G$ where $H_0$ is the Hubble 
constant.      The matter and radiation densities scale as  $\rho_m / \rho_c = \Omega_m/a^3$  and $\rhorad/\rho_c  = \Omegarad/a^4$,  where
$\Omega_m,  \Omegarad$  are the current fractions of the critical density at time $t=t_0$ where $a(t_0) =1$.    The first Friedman equation becomes, 
  when $g =1$, 
\beq
\label{F1Matter}
\frac{2}{3H_0^2}  \[   \(  \adota \)^2 -  \addota  \]  =   \frac{\Omega_m}{a^3}   +  \frac{\Omegarad}{a^4}  .
\eeq
When  $\Omegarad = 0$,  the general solution,  up to  a time translation,  is 
\beq 
\label{asolMatter}
a(t) =  \(  \frac{\Omega_m}{\mu} \)^{1/3}  \Bigl[  \sinh ( 3 \sqrt{\mu} H_0 t/2) \Bigr]^{2/3} .
\eeq
The constant $\mu$ is fixed by $a(t_0) = 1$.     One can check that $\rhovac$ is indeed constant in time:
\beq
\label{rhohatconst}
\frac{\rhovac}{\rho_c} = \mu .
\eeq
which implies that $\mu + \Omega_m =1$,  i.e. $\mu$ is just $\Omega_{\rm vac}$  
However,  when $\Omegarad \neq 0$,   $\rhovac$ is no longer constant in time.   This can be proven directly from the Friedmann equations,  or if one wishes,  numerically.  

\def\tH{t_H}
Thus,  there is a choice of vacuum  with a back-reaction that  is  entirely consistent with the current era, namely
our solution to $a(t)$ is identical to the $\Lambda$CDM model.    
At early times,   $a(t) \propto t^{2/3}$,  i.e. matter dominated,  and at later times grows exponentially, 
$a(t) \propto \exp(  \sqrt{\mu} H_0 t )$,  i.e. is dominated by vacuum energy.   
Given $\Omega_m$,  then  the equation $a(t_0) =1$ determines  $\tH \equiv H_0 t_0$ and thus the
age of the universe.     Observations indicate $\Omega_m = 0.266$,   and eq. (\ref{asolMatter}) 
 gives  $t_H = 0.997$.     The reason this is so close to the measured value of $t_H = 0.996$ is that 
 radiation is nearly negligible.    
 
\begin{figure}[htb] 
\begin{center}
\hspace{-15mm} 
\psfrag{x}{$z $}
\psfrag{y}{$\adot/a H_0$}
\includegraphics[width=7cm]{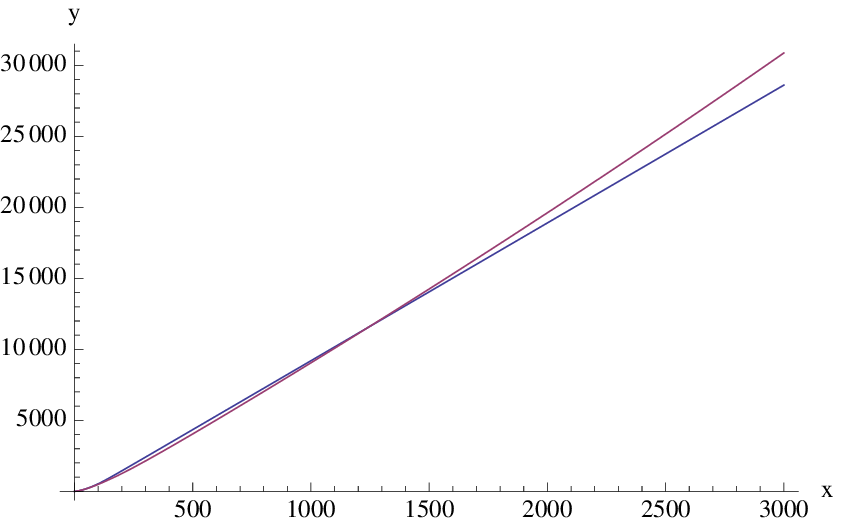} 
\end{center}
\caption{The Hubble constant as a function of redshift $z$ for the solution to our model eq. (\ref{F1Matter})  including
radiation,    
verses the  Friedman equation   eq. (\ref{Friedmann1})  with radiation,  matter and a standard cosmological constant 
with $\Omega_m =0.266,  \Omega_{\rm rad} = 8.24 \times 10^{-5},  \Omega_\Lambda = 1- \Omega_m - \Omega_{\rm rad}$.}
\vspace{-2mm}
\label{Compare} 
\end{figure}

 It is interesting to compare our model with the standard model of cosmology when one includes
 radiation,  since,  as explained above,  $\rhovac$ no longer behaves like a cosmological constant.     
 In Figure \ref{Compare}  we compare the expansion rate $H=\adot/a$ as a function of redshift $z$.
 One sees that for  redshifts $z= 1- 1/a(t)$ up to at least $1000$,  there is only a small 
  discrepancy between our model and the standard model of cosmology.    Interestingly, 
  our vacuum energy ceases to behave as a cosmological constant roughly around the time
  the Cosmic Microwave Background was formed.     
 
The condition $g=1$ relates Newton's constant $G$ to the cut-off $k_c$.  There are a number of possible
interpretations of this curious result.   Recall the Planck scale $k_p$ is simply the scale one can define from $G$,  but it is not a priori a  physicallly meaningful energy scale;   rather it is just the scale that one naively {\it expects}  some form of quantization of
the gravitational field to become important.    Here,  the relation $g=1$ is a specific relation between the 
cut-off,  Newton's constant G, and the number of particle species,  and is unrelated to the quantization of gravity itself.   One interpretation is simply that the cut-off $k_c$  is the fundamental scale  and  that $G$ is not  fundamental,
     but rather is fixed by the cut-off from $g=1$.

Allow us to speculate further:   the relation $g=1$  suggests that {\it gravity itself originates from quantum
vacuum fluctuations}.     Let us now argue how gravity can be  heuristically ``derived'' from quantum mechanics.    
Processes in a closed universe are adiabatic,  $dQ =0$,  and the first law of thermodynamics is
$dE = - p dV$,   where $p$ is the total pressure due to all constituents regardless of their nature.  
   Since $\rhovac$ is ``kinetic'' in that it depends on time derivatives,   in analogy with internal kinetic energy, 
   let  us identify the internal energy with vacuum energy,  $dE = \rhovac dV$.   
Recalling that $\rhovac$ is proportional to $\CA$,  eq.  (\ref{rhovacphys}) to lowest order, the first law  is then nothing other than the second 
Friedmann equation (\ref{Friedmann2}),   with Newton's constant identified as $G= \frac{ 8 \pi}{3 \Delta N k_c^2}$,  which is the same
as $g=1$.    
If one  assumes energy conservation $D^\mu T_{\mu \nu} =0$,  then this implies eq. 
(\ref{energyconservation}),  from which,  together with eq. (\ref{Friedmann2}),    one can derive the first Friedmann equation (\ref{Friedmann1}).   
     It is well-known that the  first Friedmann equation can be derived 
from  non-relativistic  Newtonian gravity,   so the above argument indirectly implies Newton's law of 
gravitation.   
 From this point of view,  the fundamental constants are $\hbar, c$ and $k_c$,  
and Newton's constant $G$ is emergent.   
Curiously, in a universe with more bosons than fermions,  gravity would actually  be repulsive.   The Planck scale has lost any real physical meaning here,   and gravity is very weak simply because
the cut-off $k_c$ is large.   
 If there is any truth to this idea,  it renders the goal of quantizing gravity obsolete since it is
already a quantum effect!     Gravity would then be the ultimate macroscopic quantum mechanical phenomenon.

Finally we wish to make some observations on the so-called cosmic coincidence problem.
Simply stated,  the problem is that at the present time $t_0$  the densities of matter and vacuum energy
are comparable,  and since they evolved  at different rates,   their ratio would  apparently 
have had to be fine-tuned 
to differ by many orders of magnitude in the very far  past.   
From the point of view of  the second order differential eq. (\ref{F1Matter}), $\mu = \Omega_{\rm vac}$ is just 
one of its arbitrary integration constants and we cannot predict it. 
   However our construction has a bit more to it,  based 
on the detailed equation (\ref{rhovacphys}).      First of all,  in our  approach to the 
problem,  $\rhovac$ is determined by current era physics,   and is guaranteed to be of order 1 if 
$k_c$ is close to the Planck scale;   this is how our proposal involves UV/IR mixing.   
Using $\CA \approx 9 H_0^2 /4$,  which is an observational input,   one finds    $\rhovac/ \rho_c \approx  \frac{3 \Delta N k_c^2}{8 \pi k_p^2}$.
Note also that if one rules out ``phantom energy'' with $w<-1$ based either on its strange 
cosmological properties\cite{phantom} or general  thermodynamic arguments\cite{ALthermo},   then this implies  implies $g<1$ which gives $\rhovac/\rho_c  < 1$.    

One can argue further.   
Let $t_m$ be the time beyond which radiation can be neglected.  
In the solution eq. (\ref{asolMatter}),  $t$ should be replaced by $t-t_m$.  To a good approximation, $t_0 - t_m \approx t_0$.     
Imposing then $a(t_0) =1$ in equation (\ref{asolMatter}) with $\mu +\Omega_m=1$ 
determines  $H_0 t_0$ as a function of $\Omega_m$. One can show that $2/3 < H_0 t_0 < \infty$.  
As stated above,  with the present data, $\Omega_m=0.226$, this gives $H_0 t_0=0.997$.
Thus the ``coincidence''  that $\rho_m / \rhovac \approx 0.37$  is now linked to the fact that 
current measurements give $H_0 t_0 $  very close to $1$. 
How would these numbers  change if they had  been measured in the past, 
say at time $t'=t_0/2$, nearly  7 billion years ago when the universe was half  as old?  
Using eq. (\ref{asolMatter}) with $\Omega_m = 0.266$,  one finds that 
at the time $t_0/2$   the Hubble constant was $H'=1.52\, H_0$ and $H't'=0.76$.
As we did,  an observer at that time interprets the solution with $a(t') =1$,  and from 
eq. (\ref{asolMatter}) we can infer the value of $\Omega_m'$;     
 one finds  $\Omega_m'=0.68$ and $\rho'_m/\rho'_{\rm vac}= 2.13$. 
Thus for the entire duration of the second half of the universe's history,   
the product $H't'$ only varied  by a factor $3/4$ and the ratio $\rho'_m/\rho'_{\rm vac}$ by less than  a factor  of 
$6$. Actually the evolution of this ratio is  very slow.     For instance,  if one would have measured it at time $t_N=t_0/N$,  one would have obtained  $(\rho_m/\rho_{\rm vac})(t_N)=0.61  \, N^2$ for $N$ large.  Thus  one has to go very  deep into  the past to have  a huge difference between $\rho_m$ and $\rhovac$.  Furthermore,   at these very early times the radiation plays a role and our model breaks down,   perhaps with  $\vac$ being replaced 
by $\vachat$ of the next section.  

 An alternative way of summarizing what we have added to the discussion
of the cosmic coincidence problem is the following:     if one takes the point of view that the scale of vacuum
energy $\rhovac$ is not determined by Planck scale physics,    but rather by current day physics,  as in our model, 
then there is much less of a need to explain  any fine-tuning in the very far past.   All one needs is a high
energy cut-off,  which is within the framework of low-energy quantum field theory as we currently understand it. 
In our model,   vacuum energy is a {\it low-energy,  IR}  phenomenon,  like the Casimir effect,  but is
also influenced by UV physics, via the cut-off.

\section{Vacuum energy plus radiation}  

\subsection{Vacuum energy in the conformal time vacuum} 

In this section,  we show that another choice of vacuum $\vachat$  is consistent with a universe consisting only
of vacuum energy and radiation, i.e. massless particles. 
    The quantization scheme is based on conformal time $\tau$,  defined as 
$dt =  a  d\tau$.   Like the choice in the last section,  this also simplifies the  time dependence  of the action  (\ref{Phiaction}),  and is common in the literature.   (See for instance \cite{BirrellDavies,Jacobson} and references therein.)  Rescaling the field $\Phi =  \phi /a$,  and integrating by parts,  the action becomes 
\beq
\label{phiAction}
S =  \int  d \tau  d^3 x \(   \inv{2}  \d_\tau \phi  \d_\tau \phi   -\inv{2}  \gradvec  \phi \cdot \gradvec  \phi   +  \frac{\Ricci a^2}{12} \phi^2  \) ,
\eeq
  with  the Ricci scalar $\Ricci= 6   a''/a^3$,  where primes indicate derivatives with respect to
conformal time $\tau$.  

The field can be expanded in modes 
\beq
\label{modeshat}
\phi = \int  \frac{d^3 \kvec}{(2\pi)^{3/2} }  \(  a_\kvec  \, v_\kvec \,  e^{i \kvec \cdot \xvec} 
+  a^\dagger_\kvec  v_\kvec^*  \,  e^{-i \kvec \cdot \xvec }  \) ,
\eeq
where  the $a_\kvec $'s  satisfy canonical commutation relations as before.   
 The function $v_\kvec$ is now required to 
satisfy
\beq
\label{vEq}
(\d_\tau^2 +  \omegahat_\kvec^2  )  v_\kvec = 0,      ~~~~\omegahat_\kvec^2 
   \equiv \kvec^2    -  \Ricci a^2 /6 .
\eeq 
The analysis of the last section applies with $\CA$ replaced by $\Ricci/6$,  which leads to:  
\beq
\label{rhovachat}
\rhovachat  \approx   \Index \[    \frac{ \Lambdac^2}{96\pi^2}   \Ricci    
-   \inv{4608 \pi^2}  \Ricci^2  \] .
\eeq
It is clear that  $\vac \neq \vachat$  since the  $v_\kvec \neq u_\kvec$.   

It is instructive to compare the above result  with the detailed point-splitting calculation  performed in \cite{BunchDavies} for de Sitter space.   
The regularization utilized there insists on a finite answer and thus discards the $k_c^4$ and  $k_c^2$ term:  
\beq
\label{Bunch}
\langle  T_{\mu\nu}  \rangle_{\rm ren} = -  g_{\mu\nu}  \(     \inv{128 \pi^2 } (\xi -1/6)^2 \Ricci^2  -  \inv{138240 \pi^2}  \Ricci^2  \)  
\eeq
where $\xi$ is an additional  coupling to $\Ricci \Phi $ in the original  action.   
In our calculation $\xi =0$,  and one sees that our simple calculation  reproduces the first $\Ricci^2$ term.    
 When $\xi=1/6$ the theory is conformally
invariant and the additional term  is the conformal anomaly\cite{Anomaly},  which our simple calculation has missed.
This is not surprising,  since the anomaly depends on the spin of the field,  and not simply of opposite sign
for bosons verses fermions.       In any case,   in our approximation we are dropping the
$\Ricci^2$ terms.       What this indicates is that the assumption [iv]  in the Introduction is essentially correct
if one carefully constructs the full stress tensor in a covariant manner,  such as by point-splitting.

\subsection{Consistent backreaction}

In this case define  the dimensionless constant:
\beq
\label{GLambdahat}
\ghat   =      \frac{\Index }{3 \pi}   G \,  \Lambdac^2  
\eeq
such that 
\beq
\label{rhovacghat}
\rhovachat  = \frac{\ghat}{32 \pi G}  \, \Ricci .
\eeq
Including $\rhovachat$ in $\rho$,   the first Friedmann equation then becomes
\beq
\label{Fried1hat}
\( 1-\frac{\ghat}{2}  \) \( \adota \)^2  -  \frac{ \ghat }{2}   \addota  =  \frac{8 \pi G}{3} \(  \rho_m  + \rhorad \) .
\eeq

Similarly to what was found in the last section,   a consistent solution only exists  when $\ghat =1$,  but this time
with no matter,   $\rho_m=0$.         
First consider the case where there is no radiation nor matter.     Then eq. (\ref{Fried1hat}) implies 
$\addot/a =(2-\ghat) (\adot/a)^2 /\ghat$.  
Using this,  the pressure can  again be found from eq.  (\ref{Friedmann2})
\beq
\label{pvachat}
 \pvachat  = -  \frac{(4-\ghat )}{3\ghat} \,  \rhovachat .
\eeq  
Consistency  requires  the equation of state parameter $w=-1$,  i.e.   $\ghat =1$.      
The solution is    $a(t) \propto  e^{Ht}$  for some constant $H$,  and $\rhovachat$ is independent of time.

\def\tin{t_i}

  Now,  let us include radiation.  
At a fixed time $\tin$  define  $\rho_i =  3 H^2 /8 \pi G$ where $H$ is a constant equal to 
$\adot/a$ at the time $\tin$.        Now we have to solve (when $\ghat =1$):
\beq
\label{FriedRadhat}
\inv{2H^2}  \[  \( \adota \)^2 - \addota \]  =  \frac{\Omegarad}{a^4}
\eeq
where $\Omegarad =  \rhorad/\rho_i$ at the time $\tin$ where $a(\tin)  =1$. 
The general solution, up to a  time shift, is 
\beq
\label{arad}
a(t) =  \( \frac{\Omegarad}{{\nu}} \)^{1/4}  \sqrt{ \sinh ( 2 H  \sqrt{ \nu} \, t )},
\eeq
where $\nu$ is a free parameter.   
Surprisingly,  again  $\rhovachat$ is still a constant, 
\beq
\label{rhoratio}
\frac{\rhovachat}{\rho_i} = \nu .
\eeq  
However this is spoiled if there is matter present (see below).     
At early times,  radiation dominates,   $a(t)  \propto t^{1/2}$,  and at later times 
vacuum energy dominates,  $a(t)  \propto \exp( \sqrt{\nu} H t )$.  

\begin{figure}[htb] 
\begin{center}
\hspace{-15mm} 
\psfrag{x}{$Ht $}
\psfrag{y}{$\rhovachat/\rho_i$}
\includegraphics[width=7cm]{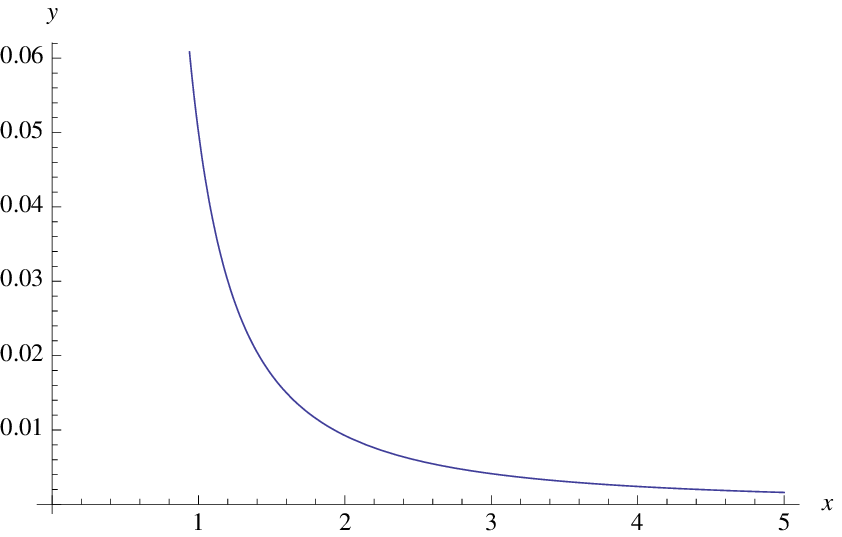} 
\end{center}
\caption{The vacuum energy $\rhovachat/ \rho_i$  as a function of $Ht$  for  $\Omegarad = 0.8$ and $\Omegam = 0.15$.}  
\vspace{-2mm}
\label{Exit} 
\end{figure}

This choice of vacuum and self-consistent back-reaction is perhaps relevant to
the very early universe which consists primarily of radiation and no matter.   
In fact,  at the very earliest times,  $H$ is presumably set by the Planck time
$t_p = 1/E_p$,  which is a much larger scale than $H_0$ by many orders of magnitude.  
In fact,  since the only scale in $\rhovac$ is $k_c$,   we expect that higher orders in
the adiabatic expansion give $H /k_c $ of order one\cite{Hkc}.  
For $k_c$ near the Planck scale,  then $H$ is roughly of the right scale for inflation.  
When vacuum energy dominates,   $a(t)$  then grows exponentially on a time scale 
consistent with the inflationary scenario\cite{Guth,Linde,Steinhardt}.    Here this is accomplished
without invoking an inflaton field.     Many  models of inflation typically suffer from
the ``graceful exit problem'',  i.e. inflation must come to an end in a relatively  short period of time.  
Based on our work,  we suggest the following scenario.   Initially there is only radiation
and vacuum energy,  which consistently leads to inflation.    However as matter is produced,
perhaps by particle creation from  the vacuum energy,  the above solution is no longer consistent. 
Thus,  $\vachat$ is no longer a consistent vacuum,  which suggests that $\rhovachat$ should somehow
relax to zero.       In support of this idea,  we numerically solved eq. (\ref{FriedRadhat}) with an 
additional matter contribution on the right hand side equal to  $\Omegam/a^3$.   As expected
$\rhovachat$ is no longer constant,   but decreases in time as shown in Figure \ref{Exit}. 
Of course,  we are already aware that eq.(\ref{FriedRadhat}) with an additional $\Omegam \not= 0$ is not consistent with eq.(\ref{Tuu}) since for a time varying $\rhovachat$ the pressure $p_{\widehat{\rm vac}}\not=-\rhovachat$;    however this plot does indeed show that $\rhovachat$ decreases.

\section{Discussion and Conclusions}

In this work we have proposed a different point of view on the Cosmological Constant Problem.   
In analogy with the Casimir effect,  we proposed the principle that empty Minkowski space should be gravitationally stable 
in order to fix  the zero point energy which is otherwise arbitrary.   
In a FLRW cosmological geometry,  this leads to a prescription for defining a physical vacuum energy $\rhovac$  which depends on $\adot$ and $\addot$.    In the current era,
this leads to a $\rhovac $ that is constant in time with $\rhovac \approx  k_c^2  H_0^2$,  which is the correct order of magnitude in comparison to
the measured value if the cut-off $k_c$ is on the order of  the Planck scale.   

    We described two different choices of vacua,  and studied the 
self-consistent back-reaction of this vacuum energy on the geometry.     One choice of vacuum is
consistent with the current matter and dark energy dominated era.   Another choice of vacuum 
is  consistent with the early universe consisting of only radiation and vacuum energy,   and we 
suggested that this perhaps describes inflation,  and also a resolution  to the graceful exit problem.  
Although  our proposals  could certainly be further improved,  their consequences  have  at least 
survived  a few checks.  
The  role of  higher orders of the adiabatic expansion on the back reaction should be better deciphered.

Both these consistent solutions require a relation between the cut-off $k_c$ and Newton's constant $G$, and we speculated above on possible interpretations of this relation.
It remains unclear how to apply the ideas of this work to the time period intermediate between inflation and the current era,  where in our scenario,  $\vachat$  would somehow evolve to $\vac$,  and this is clearly beyond the scope of this paper.

\section{Acknowledgments}
  We wish to thank the IIP in Natal,  Brazil,   for their hospitality during which time  this work was initiated.   This work is supported by the National Science Foundation under grant number  NSF-PHY-0757868 and by the ``Agence Nationale de la Recherche'' contract ANR-2010-BLANC-0414.


\begin{thebibliography}{99}
\bibitem{convert}
We will mainly work in energy units.   
We remind the reader that  an inverse length $1/L$ can be converted to an energy $1/L \to \hbar c/L$, and similarly for  an inverse time $1/t \to \hbar/t$.   The Planck energy $E_p$ and momentum $k_p$ are 
$E_p  = \hbar c  \Lambdap   =  \sqrt{\hbar c^5 /G } = 1.2 \times 10^{19}  \,{\rm  Gev}  $.  
The current value of the Hubble constant is $H_0 = 2.3 \times 10^{-18}  \,  {\rm s}^{-1} =  1.5 \times 10^{-42} ~ {\rm Gev}/\hbar$.    Unless otherwise indicated we set $\hbar = c =1$.
\bibitem{Maggiore}    After completion of this article we were made aware of some recent works 
 where similar but not identical ideas were pursued\cite{Maggiore1,Maggiore2,Maggiore3,Deiss}.   
 The first works  \cite{Maggiore1,Maggiore2,Maggiore3}  considered  $\rhovac \propto H^2$,  which is time-dependent and 
 ruled out by observations.    The later  work \cite{Maggiore3}  briefly considered   
 the conformal time vacuum $\vachat$,  described  here in section III;   based on our understanding  this
 vacuum is  also inconsistent with the present universe,  again because it is time-dependent,
 in  contrast  to our vacuum $\vac$ described in section II.
\bibitem{footnote2}  Of course, we can adopt a hydrodynamic point of view so that the cosmological dark energy fluid is locally characterized by the flat space vacuum energy. But then this vacuum energy is a constant and moduli independent,  and thus  ill-defined unless we invoke  some principle fixing the zero point energy in flat space.
\bibitem{zero1} Consider  the  action for a single bosonic field  $\phi$ in zero spatial dimensions:
$S = \int dt \( \inv{2}  \d_t \phi \d_t \phi - \frac{\omega^2}{2} \phi^2 \)$.
The equation of motion is $\d_t^2 \phi = - \omega^2 \phi$,  thus $\phi $ can be expanded as follows:
$\phi =  \(  a e^{-i\omega t}  + a^\dagger e^{i\omega t} \)/ \sqrt{2\omega}$.
Canonical quantization gives $[a, a^\dagger] = 1$,  and the hamiltonian is
$H = \inv{2} \(  \d_t\phi \d_t \phi  + \omega^2 \phi^2 \)  =  \frac{\omega}{2} ( a a^\dagger + a^\dagger a)  =    \omega ( a^\dagger a + \inv{2} )$.
\bibitem{zero2}  Consider 
a zero dimensional  version of a Majorana fermion with action
$S = \int dt  \(  i \psibar \d_t \psibar - i \psi \d_t \psi - \omega \psibar \psi \)$
The equation of motion is 
$ \d_t \psi = -i \omega \psibar$, and $ \d_t \psibar = -i \omega \psi $,    
and thus $\psi, \psibar$ have the following expansions:
$\psi  =   \inv{\sqrt{2}} \(  b e^{i \omega t} + b^\dagger e^{-i\omega t}  \)$, and  
$\psibar =  \inv{\sqrt{2}} \( - b e^{i \omega t} + b^\dagger e^{-i\omega t}  \)$. 
Canonical quantization leads to
$\{b, b^\dagger\} = 1,  b^2 = {b^\dagger}^2 =0$,  
and the hamiltonian is 
$H =  \omega \psibar \psi = \frac{\omega}{2} \(  b^\dagger b - b b^\dagger \) =  \omega ( b^\dagger b - \inv{2} )$.
\bibitem{footnote5} If such a  scenario has some truth to it,  it suggests that not very  much ``new physics''
occurs below  the Planck scale.
\bibitem{footnote3} Shifting the potential $V(\Phi)$ of a bosonic field does not modify its equation of motion but shifts its stress tensor by a term proportional to $g_{\mu\nu}$. In a hydrodynamic description (see e.g. ref.\cite{Mukhanov}), this corresponds to a shift of the vacuum energy density $\rho$ by a constant but leaving $(\rho+p)$ unchanged.
\bibitem{Hkc}  
   Above,  as in the last section, 
we have  assumed that $\Ricci$ was small enough that we could drop the $\Ricci^2$ term in $\rhovac$.   However
in the very early universe this is not necessarily the case.   Although it is unclear whether it is justified  to 
include the higher order term in the adiabatic expansion,   if one does so,  then 
the scale of $H$ is set by the cut-off, as we now explain.  
Keeping   the additonal terms in eq. (\ref{rhovacalpha}),
including the $\log$,  assuming only vacuum energy and  that this leads to de Sitter space, 
one has $( \adot/a)^2 =  \addot/a = H^2 = {\rm const.}$.    The first Friedmann equation then becomes
an  algebraic equation  for $H$ when $\ghat =1$,  with the solution $H/ k_c  = \small{ \sqrt{2}} \,   e^{-1/4}$.
\end{thebibliography}
\end{document}